# Improved, Extended, and Total Impact Factor of a Journal

Florentin Smarandache
University of New Mexico
Mathematics and Science Division
705 Gurley Ave., Gallup, NM 87301, USA

**Abstract**

In this short paper we recall the (Garfield) Impact Factor of a journal, we improve and extend it, and eventually present the Total Impact Factor that reflects the most accurate impact factor.

## 1. Introduction

The Impact Factor (IF) or Journal Impact Factor (JIF), that is used today, was proposed by Eugene Eli Garfield (1925 –2017), an American linguist and businessman, founder of the Institute of Scientific Information (ISI), Science Citation Index (SCI), and especially Journal Citation Reports (JCR) among others The Impact Factor is computed since 1975 only for the journals registered in the database of the Journal Citation Reports (see [1]).

We call it Garfield Impact Factor (GIF) in order to distinguish it from three new types of impact factors that we propose now, in order to improve, extend, and totalize the impact factors' formulas for a better accuracy of the citations of a journal published papers.

## 2. Garfield Impact Factor

Let's consider a journal *(J)* that started on year $Y_1$. We want to compute its impact factor on year $Y_2$, where $Y_1 < Y_2$, and calculation is done on the year $Y_2+1$.

The Garfield IF of journal *(J)* on year $Y_2$ is defined as follows:

$$IF_{Y_2}^{Garfield}(J) = \frac{C(Y_2, Y_2-1) + C(Y_2, Y_2-2)}{P(Y_2-1) + P(Y_2-2)}, \qquad (1)$$

where $C(Y_2, Y_2 - 1)$ means the number of citations during $Y_2$ of journal's published articles during previous year $Y_2 - 1$;

$C(Y_2, Y_2 - 2)$ is similarly the number of citations during $Y_2$ of journal's published articles during two years ago, i.e. $Y_2 - 2$;

$P(Y_2 - 1)$ and $P(Y_2 - 2)$ represent the number of journal's published articles during years $Y_2 - 1$, and respectively $Y_2 - 2$.

$IF_{Y_2}^{Garfield}(J)$ is calculated on the next year $Y_2 + 1$.

### 3. Flaws of Garfield IF

We list the following flaws:

a) The number of citations of journal's articles published in year $Y_2$ and cited in the same year $Y_2$ are missed.

b) The journal's published articles taken into consideration are only from previous two years $Y_2 - 1$ and $Y_2 - 2$, which is superficial.

### 4. Improved Impact Factor

The case a) is always omitted by $IF^{Garfield}$ that never takes into consideration the citations in the same year as the articles were published.

An improved and more accurate $IF^{Garfield}$ is:

$$IF_{Y_2}^{Improved}(J) = \frac{C(Y_2,Y_2)+C(Y_2,Y_2-1)+C(Y_2,Y_2-2)}{P(Y_2)+P(Y_2-1)+P(Y_2-2)}, \quad (2)$$

by including the citations during year $Y_2$ of journal's papers published during $Y_2$. This is, of course, computed in year $Y_2 + 1$.

### 5. Extended Impact Factor

Case b) that shows the incompleteness of Garfield IF

$$IF_{Y_2}^{Extended}(J) = \frac{\sum_{k=Y_1}^{Y_2} C(Y_2,k)}{\sum_{k=Y_1}^{Y_2} P(k)}, \quad (3)$$

where $C(Y_2, k)$ is the number of citations during the year $Y_2$ of the journal's published articles during the year $k$;

and $P(k)$ is the number of journal's published articles during the year $k$;

of course, $k \in \{Y_1, Y_1 + 1, Y_1 + 2, \dots, Y_2\}$.

#### 5.1. Distinctions between Extended Impact Factor and Garfield Impact Factor

The main *distinctions* (with respect to Garfield Impact Factor) are the following:

- $IF^{Extended}$ shows all citations during year $Y_2$ of all journal's published articles since starting year $Y_1$; while $IF^{Garfield}$ shows the citations during year $Y_2$ of only previous two years published articles, therefore $IF^{Garfield}$ is incomplete;

- $IF^{Extended}$ also includes the citations during year $Y_2$ of journal's published articles in this year $Y_2$; while $IF^{Garfield}$ misses it, so $IF^{Garfield}$ is less accurate.

## 5. Total Impact Factor

But the best (the most accurate) and complete or exact IF is the *Total Impact Factor*, defined as below.

$$IF_{Y_2}^{Total}(J) = \frac{\sum_{k=Y_1}^{Y_2} C(k, [Y_1, k])}{\sum_{k=Y_1}^{Y_2} P(k)},$$
(4)

where $C(k, [Y_1, k])$ is the number of citations during year k of all journal's published articles during years $Y_1, Y_1+1, ..., k$ altogether, where $Y_1 \leq k \leq Y_2$, and $[Y_1, k] = \{ Y_1, Y_1+1, ..., k \}$; and $P(k)$ is the number of journal's published articles during year *k*.

## 6. *Accuracy Relationship of Order*

Let's consider the relationship of order " $>_a$ ", that means "better accuracy".

Then we have:

$$IF^{Total} >_a IF^{Extended} >_a IF^{Improved} >_a IF^{Garfield}$$

## 7. Numerical Example

| Example | Journal (J) | | | | | | | | | | | | | |
|---|---|---|---|---|---|---|---|---|---|---|---|---|---|---|
| Year of Publication | 2015 | | | | | 2016 | | | | 2017 | | | 2018 | 2019 |
| Number of published articles | 20 | | | | | 40 | | | | 50 | | | 45 | 40 |
| Year of citations | 2015 | 2016 | 2017 | 2018 | 2019 | 2016 | 2017 | 2018 | 2019 | 2017 | 2018 | 2019 | 2018 | 2019 | 2019 |
| Number of citations per year | 6 | 15 | 4 | 0 | 9 | 19 | 0 | 8 | 11 | 10 | 70 | 55 | 12 | 16 | 90 |
| Total number of citations | 34 | | | | | 38 | | | | 135 | | | 28 | 90 |

We read this table on columns, for example:

- on year 2015 the journal (J) has published 20 articles; these 2015 published articles got:

6 citations on year 2015;

15 citations on year 2016;

4 citations on year 2017;

0 citations on year 2018;

and 9 citations on year 2019;

then, the total number of citations of 2015 published papers is 6+15+4+0+9=34;

and so on;

- on year 2019, the journal *(J)* published 40 articles, and they got 90 citations on the same year 2019.

*

Let's use all four impact factor formulas to compute journal's impact factors for year 2019 (that is computing on year 2020).

1) Garfield Impact Factor for year 2019:

$$IF_{2019}^{Garfield}(J) = \frac{C(2019, 2018) + C(2019, 2017)}{P(2018) + P(2017)} = \frac{16 + 55}{45 + 50} = \frac{71}{95} \simeq 0.747.$$

2) Improved Impact Factor for year 2019:

$$IF_{2019}^{Improved}(J) = \frac{C(2019, 2019) + C(2019, 2018) + C(2019, 2017)}{P(2019) + P(2018) + P(2017)} = \frac{90 + 16 + 55}{40 + 45 + 50} = \frac{161}{135} \simeq 1.193.$$

3) Extended Impact Factor for year 2019:

$$IF_{2019}^{Extended}(J)$$
$$= \frac{C(2019, 2015) + C(2019, 2016) + C(2019, 2017) + C(2019, 2018) + C(2019, 2019)}{P(2015) + P(2016) + P(2017) + P(2018) + P(2019)}$$
$$= \frac{9 + 11 + 55 + 16 + 90}{20 + 40 + 50 + 45 + 40} = \frac{181}{195} \simeq 0.928.$$

4) Total Impact Factor for year 2019:

$$IF_{2019}^{Total}(J) = \frac{34+38+135+28+90}{20+40+50+45+40} = \frac{325}{195} \simeq 1.667.$$

Therefore, according to the accuracy relationship of order $>_a$ we have:

$1.667 >_a 0.928 >_a 1.1928 >_a 0.747$.

Whence, the exact (correct, most accurate) impactor factor of journal (*J*) is equal to 1.667.

**Conclusion**

We have defined for the first time three new types of impact factors of a journal and we designed an accuracy relationship of order.

On a numerical example each type of impact factor was computed.

Upon each impact factor's formula we clearly have: The Total Impact Factor is more accurate that the Extended Impact Factor, which is more accurate than the Improved Impact Factor, which is in its turn more accurate than Garfield Impact Factor.